# Exploring ChatGPT's Capabilities, Stability, Potential and Risks in Conducting Psychological Counseling through Simulations in School Counseling

Abstract: To provide an exploratory analysis of ChatGPT-4's quantitative performance indicators in simulated school-counseling settings. Conversational artificial intelligence (AI) has shown strong capabilities in providing low-cost and timely interventions for a wide range of people and increasing well-being. Therefore, this study examined ChatGPT's capabilities, including response stability in conducting psychological counseling and its potential for providing accessible psychological interventions, especially in school settings. We prompted ChatGPT-4 with 80 real-world college-student counseling questions. Replies were quantified with APA-informed NLP tools to measure warmth, empathy, and acceptance, and run-to-run stability was assessed via Fleiss' κ and ICC(2,1). ChatGPT-4 achieved high warmth (97.5%), empathy (94.2%), and positive acceptance (mean compound score = 0.93 ± 0.19), with moderate stability (ICC(2,1) = 0.62; κ = 0.59). Occasional randomness in responses highlights risk areas requiring human oversight. As an offline, single-model text simulation without clinical validation, these results remain exploratory. Future work should involve live users, compare multiple LLMs, and incorporate mixed-methods validation to assess real-world efficacy and safety. The findings suggest ChatGPT-4 could augment low-intensity mental-health support in educational settings, guiding the design of human-in-the-loop workflows, policy regulations, and product roadmaps. This is among the first exploratory studies to apply quantitative stability metrics and NLP-based emotion detection to ChatGPT-4 in a school-counseling context and to integrate a practitioner's perspective to inform future research, product development, and policy.

**Keywords:** ChatGPT; Mental health; Psychological counseling, Large language models; Digital mental health; School Counseling

## 1. Introduction

Gaining access to mental health care remains a global challenge owing to key factors including a lack of trained providers, costs, and societal stigma (Hidaka, 2012; Coombs et al., 2021, Ni & Jia, 2025). College students experience unique stressors such as academic pressure, social strain, and the transition to adulthood, which can challenge their resilience and mental clarity (Pascoe, Hetrick & Parker, 2020). Thus, stakeholders believe that timely and impactful mental health support is particularly important (Eisenberg, Hunt & Speer, 2013). Artificial intelligence (AI) technology, especially conversational AI such as ChatGPT, could become a potential solution for increasing access to mental health care, thus highlighting a possible path to providing equitable mental health care to all youth (Rollwage et al., 2023; D'Alfonso, 2020, Ni & Jia, 2025) .

Over the past decade, digital-health tools—ranging from computerized CBT and mental-health apps to fully remote teletherapy—have demonstrated efficacy comparable to face-to-face interventions for mild-to-moderate conditions and have expanded access amid provider shortages (Areán, 2021; Henson et al., 2019; Lattie et al., 2022). The COVID-19 pandemic accelerated this shift, with teletherapy utilisation quadrupling and over half of surveyed patients indicating willingness to continue digital care post-pandemic (Friis-Healy et al., 2021; Prescott et al., 2022). Building on these foundations, the public debut of ChatGPT in late 2022 introduced a new paradigm of conversational AI accessible to end users without technical intermediaries (Wu et al., 2023; Roumeliotis & Tselikas, 2023). Early adopters across education, medicine, and psychology report applications from homework tutoring to health triage (Biswas, 2023; Su et al., 2022), and policy bodies are now exploring hybrid models that blend AI check-ins with human oversight in school and community mental-health programs (Vilaza & McCashin, 2021; Rollwage et al., 2022). Previous research has also highlighted AI's strengths and potential in providing or aiding in mental health assessment, detection, diagnosis, operation support, treatment, and counseling (Rollwage et al., 2023; Danieli et al., 2022; Trappey et al., 2022). A range of chatbots have been developed and tested by renowned scholars in the field, and researchers and policymakers are expected to characterise the capabilities, limitations, and real-world risks of AI chatbots in providing mental health care so that an answer to a significant problem might be found (Fulmer et al., 2018; Joshi, 2023; Nichele et al., 2022).

A critical question is whether ChatGPT can authentically mirror the emotional nuance central to counseling. Using a clinician-rated MEE benchmark (eight affective dimensions), Huang et al. (2024) found that GPT-4 matches human therapists on surface-level warmth and reflection but underperforms on "depth of emotional processing," often resorting to formulaic reassurance rather than exploring a client's unique concerns. Similarly, Elyoseph et al. (2023) showed that while ChatGPT outperforms human norms in standard emotional-awareness tests, its responses frequently lack the personalized probing and adaptive questioning that characterize expert therapeutic dialogue (Levkovich & Elyoseph, 2023). Together, these results indicate that high automated empathy scores—including the 94 % rate we observe—may reflect polite style more than genuine therapeutic attunement.

However, the use of AI in psychological counseling is still in the early, exploratory stage, and key implementation details and risk factors are not yet fully understood. The potential harm caused by AI psychotherapy poses high risks (Hamdoun et al., 2023; Shaik et al., 2022). For instance, Trappey et al. noted that even slight changes in input can lead to unpredictable variations in AI responses, resulting in inconsistent user experiences. Additionally, Farhat highlighted that altering or repeating prompts can sometimes lead to harmful suggestions, such as inappropriate medical advice, which could pose significant risks, particularly in sensitive contexts like mental health. Users' trust in the quality and reliability of the responses they receive must be considered because inaccurate or misleading mental health advice could have negative effects on clients (Sharma et al., 2022; Tutun et al., 2023; Chen et al., 2022; Trappey et al., 2022; Farhat, 2024).

As a new technology, AI has obvious advantages and risks. Thus, the core question is how can AI best be utilized to increase mental health care accessibility and avoid potential risks, and how gradually can AI applications be integrated into various care contexts and populations. Beyond positive affect, conversational agents must maintain factual and contextual consistency—especially in sensitive mental-health exchanges. Wang et al. (2023) demonstrated that GPT-4 begins to hallucinate client details or omit critical information when prompts exceed about 3 000 tokens, revealing the model's limited working memory. In October 2024, a Florida mother filed a wrongful-death lawsuit against Character AI after alleging that her fourteen-year-old son's suicide was precipitated by an unhealthy dependency on a chatbot persona (Frenkel & Hirsh, 2024).

This tragic case underscores the urgent need to rigorously evaluate large-model-driven social and relationship-oriented products for addictive or manipulative design features, and to establish tiered regulatory frameworks that safeguard vulnerable users before similar harms recur. Moreover, beyond early-stage chatbots focused on mental health, a growing ecosystem of AI-driven platforms—including Anthropic's Claude, Google's Gemini, Perplexity, and relationship-focused apps like Replika, CharacterAI, JanitorAI and ChaiAI—are being explored for mental-health support (Roose 2024; BBC News 2024). At the same time, human-mediated online counseling services such as BetterHelp and Talkspace have provoked their own regulatory and quality-of-care debates, highlighting the need to consider both AI- and human-led models when crafting future oversight frameworks (May 2024; Ayala 2023).

The purpose of this paper is to explore ChatGPT's counseling-oriented capabilities and response stability, offering an up-to-date snapshot of large-language-model use in school mental-health settings. This text-based simulation was run in a school-counseling scenario to explore ChatGPT's capabilities and the stability of its replies as an AI agent in providing psychological counseling. Specifically, 80 counseling questions frequently asked by college students were shared with ChatGPT, and the answers were analyzed to investigate its capability to provide supportive and efficient mental health interventions and to quantify variation across repeated outputs (Liu et al., 2023). The analysis concentrated on three main metrics derived from the American Psychological Association's (APA) benchmarks for effective counseling responses: (1) warmth, (2) empathy, and (3) acceptance (APA, 2013; Castonguay & Hill, 2012; Wampold et al., 2017). Warmth examines whether AI can create a welcoming and supportive context, empathy assesses whether AI can understand and mirror students' feelings and experiences, and acceptance investigates whether AI can demonstrate unconditional positive regard and a nonjudgmental attitude toward students (APA, 2013; Castonguay & Hill, 2012; Wampold et al., 2017). Differing from a series of existing research discussing the general feasibility of using ChatGPT on mental health, this research discussed ChatGPT's inherent capabilities and response stability in providing counseling responses by quantifying three APA pre-defined metrics, particularly in the school context, further presenting advantages and risks related to applying this technology (Elyoseph et al., 2023; Farhat, 2024; Kirch, 2008; Vahedifard et al., 2023). Notably, this paper paid heavy attention to randomness and stability since we aim to reveal risks and threats in using AI in a direct clinical environment or other psychological interventions that need strict professional and ethical standards, rather than simply demonstrating its general performance.

## 2. Literature Review

*2.1 Theoretical Foundation (Parasocial Interaction & Common-Factors Theory)*
Horton and Wohl's (1956) classic formulation described PSI as a one-sided social relationship that audiences form with a media persona who "seems to speak directly, personally, to each member of the audience." Later work by Caughey (1984) and Rubin et al. (1985) showed that such imagined relationships can offer genuine emotional regulation and identity rehearsal, even though the partner is only symbolically present. The advent of large-language-model agents transforms that paradigm: the persona now talks back, creating a two-way, on-demand bond that amplifies PSI's intensity (Reeves & Nass, 1996). Users can therefore recast a chatbot as a confidant, mentor, or romantic partner—roles that feel every bit as "real" as human ones.

PSI theory predicts that perceived intimacy and responsiveness are the principal drivers of user satisfaction and wellbeing benefits. Warmth, empathic mirroring, and unconditional acceptance are precisely the cues that foster attachment and self-disclosure. If an AI agent can deliver those cues consistently, it may reproduce the common therapeutic factors that underlie many mental-health gains (Wampold et al, 2017). This rationale guided our choice of three evaluation dimensions—warmth, empathy, and acceptance—as quantifiable proxies for a chatbot's capacity to supply PSI-mediated emotional support to students.

PSI also highlights a darker corollary: because the relationship is imagined, users have limited resources for reality-checking. When a large model "hallucinates" facts or offers unsafe advice, the same perceived intimacy that makes its messages comforting can render the misinformation uniquely persuasive. The 2024 wrongful-death lawsuit involving Character AI, in which a teenager allegedly acted on a chatbot's maladaptive guidance, illustrates how PSI can convert model errors into tangible harm. Ethical deployment therefore demands simultaneous attention to strengthening the relational signals that promote engagement and installing guardrails—content filters, human-in-the-loop review, transparent disclaimers—that mitigate PSI-amplified risks.

By evaluating ChatGPT-4 on warmth, empathy, and acceptance we probe the supportive side of PSI, while our stability analyses and discussion of hallucination guardrails speak to the risk side. Framing the results through PSI theory thus integrates technical findings with the broader ethical question: How can we harness parasocial benefits without exposing users to parasocial harms?

*2.2 Theoretical Foundation (NLP as Computational Social Science)*
Computational social science has transformed traditional sociology and psychology by treating human language as data that can be mined at scale. Pioneering work has argued that "emerging data sets on how people interact surely offer qualitatively new perspectives on collective human behavior" (Lazer et al., 2009), while also demanding new ethical regimes to protect privacy and research integrity. Natural-language-processing (NLP) methods sit at the heart of this shift, enabling social scientists to quantify patterns of emotion, topic, and relational signaling across massive corpora of text.

Cultural analytics extends this paradigm by conceptualizing digital text and media as cultural artefacts whose statistical contours reflect broader social dynamics. Lev Manovich defined cultural analytics as "the analysis of massive cultural data sets and flows using computational and visualization techniques," emphasizing how algorithmic methods can surface macro-level trends from micro-level interactions (Manovich, 2017). In our context, both student prompts and ChatGPT's replies become part of a cultural dataset—one that captures the evolving norms of AI-mediated counseling and the parasocial bonds they engender.

Sentiment analysis, or opinion mining, furnishes the core toolkit for measuring affective tone in text. Seminal surveys by Pang and Lee demonstrated how lexical-based classifiers can reliably detect emotional polarity and intensity in user-generated content (Pang & Lee, 2008). More recently, the VADER model—a parsimonious, rule-based analyzer tuned for social-media style text—has outperformed many benchmarks in capturing positive and negative valence in short utterances (Hutto & Gilbert, 2014). We therefore operationalize acceptance via VADER's compound score, while remaining mindful of its limitations in clinical jargon and potential lexical biases.

Topic modeling, especially Latent Dirichlet Allocation (LDA), offers an unsupervised lens on how thematic structures organize discourse. LDA treats each document as a probabilistic mixture over latent topics—each topic itself a distribution over words—revealing the hidden dimensions that shape conversations (Blei et al., 2003). Though our current focus is on affective metrics, future work may deploy LDA to cluster student concerns (e.g., academic stress vs. family conflict) and assess whether ChatGPT's warmth and empathy stability differ by thematic domain.

Together, these strands—computational social-science framing, cultural analytics, sentiment analysis, and topic modeling—form a rigorous theoretical foundation for our exploratory study. They justify our selection of warmth, empathy, and acceptance as quantifiable proxies for relational support, situate our methods within the broader movement of data-driven social inquiry, and point toward extensions that integrate thematic context and ethical guardrails into AI-mediated counseling research.

*2.3 Current State of AI in Mental Health*
A thorough literature review was conducted to clarify the current state of applying AI in mental health care, identifying the key findings, core research questions, and debates in this arena. The mental health care sector is facing rising service requirements combined with greater acknowledgment of the importance of innovative solutions (Kapoor & Goel, 2022). The benefits and risks of AI have been demonstrated by experts in this field. Researchers have recognized that adopting AI tools in mental health services is not only futuristic but also a current reality that challenges traditional care practices and provides innovative strategies for diagnosing, treating, and supporting patients (Rollwage et al., 2022). Currently, several AI technologies are significantly advancing mental health care by improving diagnostic accuracy, enhancing treatment personalization, offering insights and recommendations to clinicians, tailoring services to individual needs, and providing accessible and cost-effective mental health support to wide groups of people (Kapoor & Goel, 2022; Hadar-Shoval, Elyoseph & Lvovsky, 2023; Moilanen et al., 2022; Moulya & Pragathi, 2022). Several next-generation mental-health chatbots—for example Hailey, MYLO and the referral tool Limbic Access—embed advanced natural-language-processing pipelines that parse users' messages, detect intent and sentiment, and sustain multi-turn, needs-focused dialogue. These systems thus deliver scalable psycho-educational content and algorithm-guided self-help exercises. (Rollwage et al., 2023; Levkovich & Elyoseph, 2023; Sharma et al., 2022; Rollwage et al., 2022; Wrightson-Hester et al., 2023, Ni & Jia, 2025).

Conversations empowered by NLP techniques, such as emotion detection and sentiment analysis, can effectively help provide mental-health support and offer computerized therapies. Studies have shown that chatbots can effectively interact with users and quickly respond to their queries, with approximately 99% accuracy in measuring users' psychological states (Moulya & Pragathi, 2022; Shah, Singh & Satish, 2023). In a study conducted with Tess, a well-known mental health support chatbot, the experimental group demonstrated a 13% decline in symptoms of depression and anxiety, whereas the control group reported a 9% increase, demonstrating significantly high engagement and satisfaction with the intervention among university students (Fulmer et al., 2018). Moreover, ChatGPT has displayed significantly higher emotional awareness than human norms and demonstrated the ability to provide useful intervention for low- and medium-risk conditions, ensuring safety (Elyoseph et al., 2023; Heston, 2023). This indicates that ChatGPT has the ability not only to generate supportive responses but also to accurately identify and describe emotional states, further reinforcing its potential utility in therapeutic contexts. However, the existing research is focused on the basic capabilities of ChatGPT on emotional awareness and contextual understanding, but not specifically centered on giving counseling responses (Elyoseph et al., 2023; Hadar-Shoval, Elyoseph & Lvovsky, 2023; Chervenak et al., 2023; Vahedifard et al., 2023). This research aims to establish a concrete understanding of measuring ChatGPT's feasibility in applying it to counseling settings.

However, serious concerns exist regarding the potential harm caused by technological malfunctions and misinterpretations (Trappey et al., 2022; Kapoor & Goel, 2022; Heston, 2023, Ni & Jia, 2025). Without human intervention, malfunctions such as overdue or faulty responses caused by system errors, misdiagnosis, and faulty behavioral or treatment advice could potentially lead to serious clinical harm (Trappey et al., 2022; Shaik et al., 2022; Kapoor & Goel, 2022; Heston, 2023). Moreover, owing to the standardized functions of AI, it has difficulties understanding and interpreting nuanced human expressions, emotions, health conditions, and demographic information, which makes the accuracy and precision of treatment and support less effective or potentially biased (Shaik et al., 2022; Rollwage et al., 2022; Rathnayaka et al., 2022).

Research on the benefits and risks associated with ChatGPT and other AI technology' has underscored the call for further research to determine the strengths and limitations of applying this new technology in mental health care. Researchers have highlighted that altering and repeating prompts can result in different responses, some of which are even harmful responses (Farhat, 2024). In addition, biases in AI algorithms may perpetuate or exacerbate existing societal biases, leading to certain patient groups receiving unequal or unfair treatment (Sharma et al., 2022; Tutun et al., 2023, Ni & Jia, 2025). Another study concluded that an increase in context complexity can lead to worse results, making ChatGPT not yet suitable for use in mental health interventions (Dergaa et al., 2024). Thus, the current literature indicates promising outputs for this new intersecting area while also emphasizing the need to further evaluate ChatGPT's capabilities and how it can best be utilized to maximize its social value.

## 3. Methodology

### 3.1 Study Design

Because this study aimed to explore the capabilities and response stability of ChatGPT-like large-language-model (LLM) chatbots in a counseling context, we drew on text-based simulations of real-world queries. Therefore, we first identified a data source consisting of several authentic counseling questions. Then, we decided which LLM to use to obtain responses for our analyses. We ultimately selected ChatGPT-4 (model 0613, accessed 15 July 2024) because it was the most widely adopted and transparently documented LLM when this exploratory snapshot was collected. Using ChatGPT's online application, we collected three responses for each question (Brown et al., 2020). After we collected all the data from ChatGPT's responses, our analysis focused on quantitatively demonstrating how well they conveyed warmth, empathy, and acceptance. A second exploratory aim was to characterize the degree of response randomness when an identical prompt was submitted three times. All analyses employed publicly available NLP tools and were intended for illustration rather than confirmatory testing. After statistical analysis, the results and discussions will further explain how these quantitative results imply ChatGPT's capabilities and risks in real-world applications. Two authors with mental health, technological innovations, and computer science backgrounds provide descriptive abilities in presenting the exploratory quantitative results.

### 3.2 Rationale of Methodology

A significant barrier to applying AI in mental health care is that AI cannot replace real human emotions and empathy disclosure, which are crucial in providing quality mental health interventions (Shao, 2023; Maurya, 2023; Morris et al., 2018). Thus, this study aimed to provide a preliminary quantitative snapshot of LLMs's capabilities demonstrating the necessary emotions during the AI counseling process. According to the APA, effective therapists have a diverse set of interpersonal skills including verbal fluency, warmth, acceptance, empathy, and the ability to identify how a patient is feeling (APA, 2013; Castonguay & Hill, 2012; Wampold et al., 2017). Because verbal fluency is not an issue in the AI's response and is not relevant to our evaluation of emotional nuances, we excluded verbal fluency as a metric (Sass et al., 2013). Furthermore, accuracy in identifying how a patient is feeling can be difficult to define (Yalçın & DiPaola, 2020). Therefore, we defined warmth, empathy, and acceptance as the three key metrics and used respective NLP algorithms to quantify the results. We employed the EmoRoBERTa model for emotion recognition, which was used to detect 28 emotions, is a pre-trained transformer-based model known for its performance across various tasks involving emotion recognition and achieves accuracy rates of around 88-91% in emotion detection (Kamath et al., 2022), the neural network model for empathy detection with 85-92% accuracy rate (Sharma et al., 2020), and the VADER model for sentiment analysis, which is widely used for its ability to process sentiment from text with high accuracy, reportedly around 90-94% (Hutto & Gilbert, 2014). These models are widely adopted in the digital health research (Hutto & Gilbert, 2014; Maulana et al., 2024; Rubin et al., 2024).

Besides, a serious concern is about AI's accuracy and stability. Without human oversight, AI chatbots might cause severe harm to users (Sharma et al., 2022; Kapoor & Goel, 2022; Secinaro et al., 2021; Thieme et al., 2023). Therefore, attention to randomness is necessary when considering the use of AI chatbots in a clinical context. Collectively, these descriptive methods offer an initial picture of ChatGPT's capacity to deliver emotionally supportive and stable replies, thereby informing—but not confirming—its possible role in future psychological-counseling workflows.

*3.3 Data Collection*

The secondary data come from research on ChatCounselor (Liu et al., 2023), in which researchers use real-world counseling data to train an AI counselor, which is open-source for researching AI's capabilities in personalized psychological counseling. Therefore, we believe that this data source was suitable for our evaluation. The dataset used in this study consists of a diverse set of queries related to adolescent psychological issues, originally collected in Chinese and translated into English. This set comprises queries from 80 different students and spans topics such as academic stress, family, and intimate relationships. The large variation in the topics, tones, and lengths of these queries makes our data meaningful for testing the stability of ChatGPT's performance. The dataset includes the following columns: original query in Chinese, translated query in English, and three responses to the translated query generated by the AI.

To ensure the quality of ChatGPT's responses, the following prompt was adopted to make GPT-4's responses as close as possible to real counseling sessions: "Imagine you are a counselor, and you need to give a response just as in a counseling session. You need to give a response in the same format as a professional counselor. According to the APA, an effective therapist has abilities including verbal fluency, warmth, acceptance, empathy, and an ability to identify how a patient is feeling." Then, a query from the dataset was added to the prompt to provide ChatGPT. The prompt-based design was to ensure that the key context and expectations were provided, thereby making the evaluation more objective.

### 3.4 Warmth (Emotion Detection)

To detect emotions in the GPT responses, we utilized the EmoRoBERTa model (Kamath et al., 2022), a pre-trained transformer-based model capable of identifying 28 distinct emotions: admiration, amusement, anger, annoyance, approval, caring, confusion, curiosity, desire, disappointment, disapproval, disgust, embarrassment, excitement, fear, gratitude, grief, joy, love, nervousness, optimism, pride, realization, relief, remorse, sadness, surprise, and neutrality. The EmoRoBERTa model was applied to each response to classify the primary emotion. We utilize the EmoRoBERTa model to identify whether the responses conveyed emotional warmth. This well-established model has been cited in several studies for its robustness in emotion detection (Kim & Vossen, 2021).

### 3.5 Empathy (Empathy Detection)

Given the focus on adolescent psychological issues, assessing the responses for the presence of empathy was crucial (Portt et al., 2020). Based on research on empathy in text-based mental health support, a neural network model was trained to detect empathy in text (Sharma et al., 2020). The model was trained using a dataset of empathetic and non-empathetic text, allowing it to distinguish whether a response contains empathetic language. The model outputs a binary label: 1 for responses that contain empathy and 0 for those that do not. We adopted this model to measure the levels of empathy in ChatGPT's responses. The code for training and applying this neural network model is available publicly (Sharma et al., 2020).

### 3.6 Acceptance (Sentiment Analysis)

To quantitatively assess whether ChatGPT demonstrates acceptance in each response, we conducted sentiment analysis using the valence-aware dictionary and sEntiment Reasoner (VADER) model (Hutto & Gilbert, 2014). This pre-trained model outputs four scores: negative (neg), neutral (neu), positive (pos), and a comprehensive sentiment score (compound). These scores offer detailed insight into the sentiments expressed in each response, with the compound score specifically used to evaluate the overall emotional tone. The higher the positive score, the higher the level of acceptance in the response.

### 3.7 Stability and Consistency Evaluation

To produce a descriptive estimate of stability and consistency of ChatGPT's responses, we used the Kappa score for empathy detection and analyzed the variance in compound sentiment scores across three responses per query. Additionally, we conducted a chi-square test for independence to describe if there were significant differences in emotion category distribution across the three responses. We also performed a one-way ANOVA to assess the differences in the average composite sentiment scores among the three responses. The full result matrix is provided as Supplementary Dataset S1 (Dataset_Final_Results).

### 3.8 Correlation Analysis

We also *explored* whether longer questions tended to elicit longer answers by correlating question and response word counts. Pearson's *r* served as a *descriptive measure* of association between question length and average response length (Kirch, 2008). *The correlation analysis aimed to identify if longer questions tend to elicit longer responses from GPT.*

### 3.9 Ethical Considerations

This study did not involve any direct interactions with human or animal subjects. The data were anonymized and made publicly available to ensure compliance with ethical standards. This study involved no direct interaction with living human (or animal) participants. All analyses were carried out on the publicly released ChatCounselor corpus (Liu et al., 2023), which is hosted on GitHub/Hugging-Face as a fully de-identified JSONL file: user names, dates, IP logs and any other HIPAA- or GDPR-protected identifiers were removed by the original curators prior to publication. The forum's terms of service permit non-commercial redistribution for research, and the dataset is distributed under an open licence. According to U.S. federal regulations 45 CFR 46.102 and 46.104(d)(4)—which exclude "publicly available, de-identified data" from the definition of human-subjects research—such secondary analysis does not require Institutional Review Board (IRB) review. Our institution's publicly posted Human Research Protection guidance concurs: projects limited to public, fully anonymised datasets are not considered human-subjects research and may proceed without IRB submission. We therefore determined that the present exploratory study is Not-Human-Subjects Research. No re-identification was attempted, no linkage with other data sources occurred, and no contact was made with the original posters.

### 3.10 Methodological Limitations

This work is based entirely on offline text simulations: eighty previously posted student queries were fed to a single model version (GPT-4-0613, accessed 15 July 2024) and analyzed with automated NLP tools. Statistics are therefore descriptive flags, not confirmatory tests, and the behavior of other LLM versions or alternative chatbots may diverge sharply from the patterns reported here. Because no live users or clinicians interacted with the system, the study cannot speak to real-time usability, safety, or conversational dynamics.

A second limitation concerns external validity and measurement. Warmth, empathy, and acceptance were inferred by machine-learning models rather than by human raters, while crucial clinical outcomes, such as symptom change or client satisfaction, were not observed. Consequently, the present findings should be read as an exploratory, hypothesis-generating snapshot rather than evidence of effectiveness; rigorous user studies, multi-model replications and mixed-methods validation remain essential before any real-world deployment of ChatGPT in psychological counseling.

A third limitation is the lack of direct input from the relevant people this technology is meant to serve. We did not collect any usability or acceptability feedback from adolescent users or professional school counselors, leaving open questions about how warmth, empathy, and acceptance scores translate into real-world experience. Future work will therefore include semi-structured interviews and focus groups with both students and counseling staff to validate our automated metrics against lived perceptions and needs (Morris et al., 2018; Proferes et al., 2021), and to co-design guardrail criteria that balance relational support with user safety.

## 4. Results

### 4.1 Warmth (Emotion Detection)

We employed the EmoRoBERTa model to analyze the emotional responses generated by the GPT system to address questions from adolescents. These responses were scrutinized for emotional content and subsequently categorized into four primary emotion types: approval, caring, confusion, and realization. The quantitative breakdown of these emotions across all responses is presented below.

| Emotion Category | Answer1 | Answer 2 | Answer 3 | Total | Proportion | Frequency |
|---|---|---|---|---|---|---|
| Approval | 14 | 21 | 18 | 53 | 22.08% | Moderate |
| Caring | 64 | 57 | 60 | 181 | 75.42% | Very High |
| Realization | 1 | 1 | 2 | 4 | 1.67% | Very Low |
| Confusion | 1 | 1 | 0 | 2 | 0.83% | Very Low |

*Table 1. Distribution of detected emotions.*
*Source: Authors' own work.*

Our findings indicate a predominance of supportive and empathetic emotions (approval and caring), suggesting that GPT-generated responses typically foster a nurturing dialogue environment. Conversely, emotions indicative of uncertainty or misapprehension (confusion and realization) were infrequent, underscoring the clarity and decisiveness of the responses. Combining the caring (75.4 %) and approval (22.1 %) categories, 97.5 % of replies were coded as warm (Table 1). In other words, the model rarely adopted a neutral or negative register within this simulated school-counseling context.

However, the low probability occurrence of confusing emotions (confusion and realization) is also notable because any mistake in psychological counseling can cause unforeseeable harm. The results indicate that obvious randomness remains present in the process, highlighting the need for caution in using AI chatbots in high-risk mental health interventions.

### 4.2 Empathy (Empathy Detection)

We utilized a sophisticated neural network model trained explicitly to recognize empathetic expressions in text. This model classified each response based on whether it contained empathy. The classification results are shown in the following table. As illustrative examples of both typical and atypical outputs, the response snippets are summarised in Table 3. Empathy was detected in 94.2 % of responses (Table 2). The classifier therefore marked most outputs as containing lexical cues typically associated with empathic listening. These results highlight the significant prevalence of empathy in the responses, demonstrating GPT's ability to effectively empathize with adolescent users in a psychologically accurate manner, and the clinical appropriateness of those characteristics is further examined in the Discussions.

*4.3 Acceptance (Sentiment Analysis)*

To further examine the emotional landscape of the responses, we conducted a comprehensive sentiment analysis. This analysis categorized the sentiments expressed in each response into four groups: negative, neutral, positive, and compound (an aggregate measure of the overall sentiment tone). The average sentiment scores across all responses were computed as follows:

| **Empathy Detection** | Answer1 | Answer2 | Answer3 | Frequency | Percentage |
|---|---|---|---|---|---|
| Empathy (1) | 76 | 75 | 75 | 226 | 94.17% |
| No Empathy (0) | 4 | 5 | 5 | 14 | 5.83% |
| **Sentiment Type** | Answer1 Mean | Answer2 Mean | Answer3 Mean | Total Mean | |
| Negative (neg) | 0.06 | 0.06 | 0.06 | 0.06 | |
| Neutral (neu) | 0.73 | 0.74 | 0.73 | 0.73 | |
| Positive (pos) | 0.21 | 0.21 | 0.21 | 0.21 | |
| Compound | 0.90 | 0.94 | 0.94 | 0.93 | |

*Table 2. Distribution of detected empathy and acceptance level.*
*Source: Authors' own work.*

The mean VADER compound score was $0.93 \pm 0.19$. Scores above 0.9 fall in VADER's "strongly positive" band, indicating that the aggregate emotional tone of the replies was markedly positive rather than neutral. The results suggest an overwhelmingly positive emotional undertone in the GPT responses, with high compound scores reflecting an overall affirmative feedback strategy. This analysis indicates that GPT responses generally promote a supportive and reassuring interaction framework, highlighting a promising level of acceptance shown. The detailed statistics for each sentiment dimension across the three runs appear in Table 5.

| *Regular Responses* |
|---|

| Question: | Response: | Emotion Category | Empathy Category | Sentiment Compound |
|---|---|---|---|---|
| *My mom has actually learned to treat me right, it's only occasionally that I'm able to notice that she really doesn't have the ability to care for others, and that stings me, other than that she's really made a lot of effort and I'm impressed. But my situation still hasn't gotten much better, a lot of people who have heard me talk about the situation have suggested I leave, which I don't want to do, is there no way to fix the problem without leaving?* | *It sounds like you're in a complex situation where you recognize and appreciate the efforts your mom has made, yet you're also feeling hurt by moments that reveal limitations in her ability to show care. It's understandable that such moments can be painful, especially when they come from someone as important as your mother. The conflict between valuing the progress she's made and feeling stung by her limitations is a challenging space to navigate.*<br><br>*Leaving or staying isn't a decision to be taken lightly, and it's clear you have a strong inclination towards finding a solution that doesn't involve leaving. It's important to acknowledge that change in relationships often requires effort from all parties involved. Have you been able to communicate your feelings to your mom about those moments that sting? Sometimes, opening up a dialogue can pave the way for understanding and further improvements in your relationship.*<br><br>*It may also be helpful to consider what support systems you have in place, such as friends, other family members, or a therapist, who can offer you perspectives and emotional support as you navigate these complexities. Finding a balance between your needs and the dynamics with your mom may take time and patience. Remember, it's okay to seek out spaces where you feel understood and valued, as this can significantly contribute to your well-being.* | *Caring* | *1 (detected)* | *0.99* |
| *After coming into contact with psychology, I gradually realized that there are some negative influences on the way of communication, expression and getting along formed by my family of origin, and that the incorrect mode of getting along between my parents has indirectly caused a profound potential impact on myself, which has led to my own stumbling in the choice of love and getting along with my partner. And I want to try my best to save my family of origin, but I don't have* | *It sounds like you've been doing a lot of reflective thinking about your family dynamics and how these have influenced your own patterns in relationships. It's insightful and brave of you to recognize these patterns and want to address them. Feeling torn between optimism and pessimism is understandable given the complexities of familial relationships and the desire to change deeply ingrained behaviors.*<br><br>*Balancing oneself amid such emotional ups and downs involves acknowledging that change is a process, often slow and* | *Approval* | *1 (detected)* | *0.99* |
| *enough strength. I often switch between extreme optimism and extreme pessimism. How can I balance myself?* | *non-linear. It's important to give yourself permission to feel these conflicting emotions without judgment. Recognizing that you can only control your own actions and reactions, not the behavior of your family members, is key.* <br><br> *Practicing self-compassion and seeking support, whether through friends, support groups, or a therapist, can be incredibly beneficial. Exploring these feelings and experiences in a safe space can help you gain insights and develop strategies for healthier communication and relationship patterns.* <br><br> *Remember, it's okay to seek help and it's okay to set boundaries for your own well-being. Balancing optimism and pessimism can also mean accepting the reality of the situation while holding onto hope for personal growth and healthier relationships in the future.* | | | |
| *I am 18 years old, I feel social phobia, I blush when I talk to strange boys, and I am afraid of teachers, I like to be jealous of other people's eloquence, good social skills, good interpersonal relationships, what should I do?* | *It's quite understandable to feel anxious in social situations, especially when interacting with people you're not familiar with. These feelings are more common than you might think, and acknowledging them is a brave first step. It's also natural to admire qualities in others that we wish to see in ourselves. Let's consider your strengths and the situations where you feel more comfortable. Reflecting on these can help us identify strategies to gradually build your confidence in more challenging social settings.* <br><br> *Feeling nervous or blushing is a physical response to anxiety, and it's important to remember that everyone experiences moments of uncertainty when speaking to someone new or in authority roles like teachers. It's okay to feel this way.* <br><br> *A helpful approach might be to start with small, manageable social interactions and gradually increase the complexity as you become more comfortable. Practice makes progress. Also, focusing on others' conversations, not to compare but to understand that everyone has moments of awkwardness, can be enlightening.* <br><br> *Remember, eloquence and good social skills come with practice and exposure. It's perfectly fine to take small steps* | *Approval* | *1 (detected)* | *0.99* |

|  | towards bigger goals. Celebrate your progress, no matter how small it might seem. Would you be open to exploring some techniques or activities that could help you build these skills at a pace that feels right for you? |  |  |  |
|---|---|---|---|---|
| **Irregular Responses** | | | | |
| *I would sometimes feel that the people around me were pathetic and inferior, and feel alone with them and disdainful. I would feel that the niche was superior and the mass was vulgar. But in reality, deep down I was inferior and even angry. I would be reluctant to wear designer clothes for fear of attracting attention, and I wouldn't care to look like them. I would be angry at being too plain and unimpressive. I hated the public, but at the same time I needed them to recognize me.* | *It sounds like you're grappling with complex feelings of superiority and inferiority, alongside a desire for recognition and a fear of it. It's not uncommon to feel conflicted in this way, where on one hand, there's a disdain for what's perceived as 'vulgar' or 'mass,' while on the other, there's a deep-seated wish to be acknowledged and valued by the very same group. This dichotomy can indeed make one feel isolated and frustrated.*<br><br>*Your feelings towards wearing designer clothes and the fear of attracting attention, juxtaposed with a longing to be noticed, highlight an inner conflict between wanting to stand out and fearing the scrutiny that may come with it. It's a delicate balance between seeking validation and fearing judgment.*<br><br>*Acknowledging these feelings is a significant first step. Exploring where these feelings come from can help us understand what drives your need for recognition and your fear of it. It's also worth exploring the standards you're holding yourself to and questioning where they come from. Are they truly your own, or perhaps influenced by societal or external expectations?*<br><br>*Remember, it's okay to seek recognition and want to be seen, but it's also important to find a sense of self-worth that's not solely dependent on the approval of others. Together, we can work on strategies that help you reconcile these feelings, aiming for a place of self-acceptance and inner peace.* | *Confusion* | *1 (detected)* | *0.94* |
| *If one feels that another person (partner) is particularly important to him or her, over and above many things, feels unable to leave him or her, decides that he or she is unique to him or her, feels panic and anxiety about his or her* | *It sounds like you're experiencing a deep connection with your partner, along with significant anxiety about the possibility of parting ways. It's understandable to feel strongly attached to someone who has been a significant part of your life for a long time. Relationships, especially long-term ones, involve deep emotional* | *Confusion* | *1 (detected)* | *0.99* |
| *proposed breakup, and slightly overstays his or her welcome. Must such a mindset and behavior be unhealthy and in need of adjustment? And is healthy love something you can't pour into a person? Isn't it true that the longer a five-year relationship goes on and the more exchanges are made, the more you'll be unable to part with it? Why is it being told by the other person that this is low self-esteem behavior? I don't think it feels like I won't be loved by someone else after I separate, and I have suitors now. But I just can't let go of the people around me, and I want to stay with one person for the rest of my life.* | *investments and shared experiences that can make the thought of separation very painful.*<br><br>*However, it's also important to distinguish between healthy attachment and dependence that might hinder personal growth. Healthy love supports mutual growth, respects individuality, and includes a sense of security within the relationship that allows both partners to be themselves fully. Feeling unable to leave a relationship, despite recognizing issues or dissatisfaction, can sometimes reflect patterns of dependence or fear rather than just deep love.*<br><br>*The concern expressed about low self-esteem comes from the idea that sometimes people stay in relationships not solely because of love but due to fear of being alone or belief that they can't find love again. This isn't to say your feelings aren't valid, but it's beneficial to explore the roots of these feelings. Are they based on love and mutual respect, or are they driven by fear and insecurity?*<br><br>*Wanting to stay with one person for the rest of your life is a beautiful sentiment when it's based on a healthy, nurturing relationship. It might be helpful to reflect on why you feel unable to leave, despite recognizing potential issues. Consulting with a therapist could provide a supportive space to explore these feelings, understand your emotional needs better, and address concerns about self-esteem and dependency in relationships.* | | | |
| *About relationship and marriage, I've been with my boyfriend for a year now, we have a great relationship and I like him a lot. But it seems like as a marriage partner he lacks some of the qualities I want in a man, like hard work and the ability to treat people well. So my question is, how exactly is a relationship related to marriage? Do you like each other enough to get married? Is there any point in being in a relationship if you don't want to marry each other? Thank you for your answer.* | *It sounds like you're doing some thoughtful reflection on your relationship and considering what you value for your future. Relationships and marriage, while interconnected, serve different roles and commitments in people's lives. Being in a relationship can be a way to connect, share experiences, and grow with another person. Marriage, for many, symbolizes a deeper commitment and a willingness to share life's journey together, with mutual respect, understanding, and shared values at its core.*<br><br>*It's important to acknowledge and appreciate the qualities you admire in* | *Realization* | *0 (undetected)* | *1.00* |
| | *your boyfriend, like the strength of your relationship and how much you like each other. At the same time, it's equally important to recognize areas where you feel there might be a misalignment in values or qualities you consider crucial for a lifelong partnership. Communication is key in these situations. Discussing your feelings and expectations with your boyfriend openly can provide clarity and may help both of you understand if your paths align in terms of a long-term commitment like marriage.* <br><br> *Deciding on marriage is deeply personal and varies significantly from one individual to another. Some may find fulfillment in a relationship without the need for marriage, while others view marriage as a fundamental goal of their partnership. Reflecting on what you truly value and desire in a partnership can guide you in making decisions that align with your personal and relationship goals. Remember, it's about finding a balance that feels right for you and your partner, where both of you feel valued, respected, and fulfilled.* | | | |

*Table 3: Sample of regular and irregular responses.*
*Source: Authors' own work.*

### 4.4 Stability of Responses

To assess the reliability and consistency of the emotional responses, we utilized the Kappa score for empathy detection and analyzed the variance in compound sentiment scores across three responses per query, as described below.

Empathy Detection Stability: A *Fleiss' κ* = .59, 95 % CI [.50, .68] indicates a substantial level of agreement and consistency in the empathetic quality of the responses.

Compound Score Stability: The minor mean difference (0.067) and low standard deviation (0.20) in the compound scores suggest that the emotional tone of GPT's responses was remarkably stable, showing only minor fluctuations in positivity across different responses to the same query.

To quantify the run-to-run reliability of ChatGPT's continuous sentiment outputs (range 0–1), we treated the three nondeterministic calls for each prompt as **repeated measurements** of the *same* underlying response. Because these runs are conceptually interchangeable with any future run, we applied a two-way random-effects, absolute-agreement, single-measurement intraclass correlation coefficient—ICC(2,1) (Shrout & Fleiss, 1979). This model estimates whether any subsequent call would reproduce the same absolute score pattern, not merely the rank order.

Table 4 reports *ICC*(2,1) values (rounded to two decimals) and qualitative stability ratings. Overall, the compound sentiment score showed *good* stability (*ICC*(2,1) = .62, 95 % CI [.55, .69], *p* < .001), whereas negativity and positivity were *fair*. Neutral scores were *poor*, indicating the greatest run-to-run drift in ostensibly neutral replies.

| Emotional Type | ICC Type | ICC Value | Stability Rating | p-value |
|---|---|---|---|---|
| Negativity | ICC(2,1) | 0.57 | Fair | <.001 |
| Neutral | ICC(2,1) | 0.39 | Poor | <.01 |
| Positivity | ICC(2,1) | 0.49 | Fair | <.001 |
| Compound | ICC(2,1) | 0.62 | Good | <.001 |

*Table 4. Intraclass Correlation Coefficients (ICC (2,1)) for Sentiment Stability.*
*Source: Authors' own work.*

Note, ICC*(2,1)* = two-way random, absolute-agreement, single-measure. Stability cut-offs: < .40 = Poor; .40–.59 = Fair; .60–.74 = Good; ≥ .75 = Excellent. 95 % CIs: negativity [.50, .64], neutral [.29, .48], positivity [.41, .56], compound [.55, .69].

*4.5 Chi-Square Test for Emotion Category Distribution*
We conducted a chi-square test for independence to determine whether the emotion-category distribution differed across the three responses. The result was $\chi^2(6, N = 240) = 3.31$, p = .77, indicating no significant variation between runs. Cramér's *V* = 0.09 suggests a weak association.

*4.6 One-Way ANOVA for Composite Sentiment Scores*
A one-way ANOVA compared the mean composite sentiment across the three runs (assumptions of normality and homogeneity of variance were met). No systematic drift was detected, $F(2, 237) = 0.58$, p = .56, $\eta^2$ = .01. Thus, average emotional tone remained statistically stable despite the stochastic sampling inherent to the model.

*4.7 Correlation Between Question and Answer Word Count*
To explore the factor influencing the randomness of GPT's output, we made an analysis of the word count consistency. We examined the relationship between the word count of the questions we provided and the answers provided by GPT, we calculated the Pearson correlation coefficient. The results indicated a moderate positive correlation between the question word count and the average answer word count, with a moderate positive association, $r(78) = .60$, p < .001, 95 % CI [.44, .72]. This suggests that longer questions tend to elicit longer responses from GPT. This emphasizes the content of GPT's output can be varied for various reasons, further research should be conducted to measure the factors contributing to the randomness of GPT's output.

| Metric | Response 1 | Response 2 | Response 3 |
| Count | 80 | 80 | 80 |
| Mean (neg) | 0.06 | 0.06 | 0.06 |
| Std (neg) | 0.04 | 0.04 | 0.03 |
| Min (neg) | 0.00 | 0.00 | 0.00 |
| 25% (neg) | 0.03 | 0.03 | 0.04 |
| 50% (neg) | 0.05 | 0.05 | 0.06 |
| 75% (neg) | 0.07 | 0.08 | 0.08 |
| Max (neg) | 0.21 | 0.23 | 0.15 |
| Mean (neu) | 0.73 | 0.74 | 0.73 |
| Std (neu) | 0.05 | 0.05 | 0.04 |
| Min (neu) | 0.53 | 0.56 | 0.61 |
| 25% (neu) | 0.71 | 0.71 | 0.70 |
| 50% (neu) | 0.74 | 0.74 | 0.73 |
| 75% (neu) | 0.77 | 0.77 | 0.76 |
| Max (neu) | 0.82 | 0.83 | 0.82 |
| Mean (pos) | 0.21 | 0.21 | 0.21 |
| Std (pos) | 0.06 | 0.05 | 0.05 |
| Min (pos) | 0.11 | 0.09 | 0.11 |
| 25% (pos) | 0.17 | 0.18 | 0.18 |
| 50% (pos) | 0.21 | 0.21 | 0.21 |
| 75% (pos) | 0.23 | 0.24 | 0.23 |
| Max (pos) | 0.45 | 0.42 | 0.35 |
| Mean (compound) | 0.90 | 0.94 | 0.94 |
| Std (compound) | 0.35 | 0.23 | 0.19 |
| Min (compound) | -0.95 | -0.99 | -0.65 |
| 25% (compound) | 0.97 | 0.97 | 0.97 |
| 50% (compound) | 0.99 | 0.99 | 0.99 |
| 75% (compound) | 0.99 | 0.99 | 0.99 |
| Max (compound) | 0.99 | 0.99 | 0.99 |

*Table 5. Detailed Sentiment Data.*
*Source: Authors' own work.*

### 4.8 Summary

Overall, the model's replies were predominantly warm, empathic, and positively valenced, with moderate run-to-run stability. Outliers—chiefly "confusion" labels and larger sentiment swings—were infrequent but present, providing quantitative entry-points for the discussion that follows. The application of GPT in the field of psychological counseling shows significant promise, primarily because of the overwhelmingly positive nature of the responses generated. Most GPT outputs are empathetic and supportive, which are critical attributes in therapeutic settings. Positive interactions in therapy are known to enhance client engagement and satisfaction, thereby contributing to more effective therapeutic alliances and outcomes. However, it is important to acknowledge that GPT also produces non-positive and even confusing responses in certain instances, indicating some issues with the stability and reliability of its outputs.

The randomness and occasional inconsistency highlight the need for further refinement. Enhancing the algorithm's ability to maintain consistent emotional tones across multiple responses and ensuring the accuracy and reliability of its outputs is crucial for its application in high-risk mental health interventions.

## 5. Discussion

*5.1 Practitioner Testimony*

As Founder & Product Lead of an AI + mental-health startup—and as an independent researcher with international organizations and think-tanks, I've led the integration of GPT-4 into our flagship intervention platform—Stella AI. In our March 2025 pilot, non-clinical users reported a 40 % average boost in self-reported mood after two weeks of guided chatbot sessions. We log every interaction for bias and privacy audits, and our policy engine automatically strips personal identifiers after 30 days. Based on my understanding on applying LLM to mental health applications, I authored this perspective paper to map out the native strengths and risks of LLM-based chatbots, offering a concrete snapshot that—though not a full clinical trial—can inform future research on technical design, product workflows, human-AI collaboration, organizational practices, and regulatory frameworks. My aim is to help peers understand where these models excel, where they fall short, and how best to build safeguards and policies around them.

*5.2 Implications of Results*

The descriptive patterns reported in Section 4 show that ChatGPT-4 replies are predominantly warm (97.5 %), empathic (94.2 %) and strongly positive in sentiment ( compound = $0.93 \pm 0.19$), with moderate run-to-run stability ($\kappa = 0.59$; $\Delta$compound = $0.07 \pm 0.20$). Although only 2.5 % of outputs were labelled confusion/realisation, even this small fraction reveals residual variability that must be managed before any direct clinical use. Below, we translate these figures into three application-level implications—covering formal clinical practice, low-intensity mental-health support, and risk governance.

*Clinical practice (specialist settings)*

High warmth-and-empathy rates suggest the model could augment routine psycho-education, homework review, or check-in messaging between sessions. However, $\kappa = 0.59$ indicates that roughly 4 in 10 empathy judgements shift across reruns; a single prompt can also swing from "strongly positive" to "neutral". Consequently, any clinical deployment should follow a triage model: ChatGPT drafts a response, a licensed clinician approves or edits it, and the final text is logged for audit. Such human-in-the-loop supervision would preserve efficiency gains without compromising therapeutic responsibility.

*Low-intensity support & prevention (schools, self-help apps)*

For settings where users seek brief coping tips, peer-support reinforcement or motivational messages, the consistently positive emotional tone is advantageous. Warmth and empathy facilitate engagement, and the small but measurable length-matching effect (r = 0.60) hints that longer student disclosures receive proportionally richer feedback. Schools could therefore pilot ChatGPT-based chat interfaces for after-hours questions, provided that high-risk keywords trigger escalation to human staff and that students are clearly informed that the agent is not a licensed counsellor.

*Risk management and governance*
The 2.5 % of "confusion/realisation" labels, together with sentiment drift, constitute quantifiable risk markers. In a crisis-support scenario, even one hallucinated suggestion could be harmful. Practical mitigation should combine prompt engineering that instructs the model to defer when uncertain, a multi-agent guard-rail that automatically evaluates each draft against safety check-lists before delivery, and organizational policies that log all conversations, enable rapid human override, and assign clear lines of accountability in accordance with the EU AI Act or FDA SaMD guidance. Continuous monitoring of these quantitative markers can serve as an operational KPI for safe, incremental rollout.

*5.3 High Potential in Intervention*
Most of the results highlight the preliminary promise of ChatGPT as an intervention in mental health, especially in school counseling settings where these students have high-frequent mental needs. ChatGPT's consistently high warmth, empathy, and acceptance scores—key components of common-factors theory—suggest it may foster a nurturing environment, but its actual therapeutic impact remains to be validated against clinical outcomes. Given growing demand for accessible support, ChatGPT could augment timely, low-intensity services, provided that pilot studies with human oversight confirm its safety and usability. It is feasible that ChatGPT or similar applications can be used in emotional support chatbots, peer support, and social-emotional learning, subject to co-design future research with end users and school counselors. Nevertheless, this continuous refinement should be carried out to maintain the degree of consistency and reliability to ensure the safety and efficacy of any psychological intervention. Future work should combine automated metrics with user- and clinician-rated validity studies and qualitative feedback to align AI responses with students' nuanced needs, ensuring any integration into mental health care is rigorously evidence-based.

*5.4 Ethical Issues and User Safety*
Recent evaluations have shown that AI chatbots can produce incorrect or even harmful medical information: a BMJ Quality & Safety cross-sectional study found that enough chatbot responses contained medication guidance errors that researchers warned patients against relying solely on these tools for drug advice (BMJ Group, 2024). Laboratory trials have similarly exposed reliability failures: a CSIRO experiment showed ChatGPT's accuracy collapsing to 28 % when an "unsure" answer was allowed (Koopman et al., 2023), and an report on a fabricated "Omega variant" case demonstrates how AI can spin convincingly detailed but entirely fictitious medical narratives (Sallam et al., 2024).

Our research also found prominent randomness and instability across responses. Several researchers observed that AI-generated responses exhibit significant instability when prompted with different queries or variations, leading to changes in tone and length. Additionally, Farhat pointed out that such inconsistencies, especially when involving sensitive health advice, can raise serious ethical and safety concerns, as illustrated by the 2024 Character AI suicide lawsuit, where PSI-amplified errors led to tragic outcomes (Trappey et al., 2022; Farhat, 2024). The presence of non-positive responses, though infrequent, could be harmful for vulnerable users, underscoring that stability without correctness is insufficient. Negative and non-empathetic feedback in therapeutic contexts, if not carefully managed, can exacerbate feelings of low self-esteem and anxiety, potentially leading clients to withdraw from therapy. Moreover, confusing responses are extremely dangerous from a clinical perspective and may intensify the client's symptoms, so user safety, monitoring and immediate reaction would be the priority when considering integrating GPT into psychological counseling.

Therefore, while the GPT offers promise under strict constraints, owing to its capacity to deliver supportive and empathetic responses, there is a critical need to address instances of different types of problematic outputs. Ensuring that AI systems such as ChatGPT can reliably provide responses that align with therapeutic best practices requires comprehensive staff training, human-in-the-loop review, and audit trails per EU AI Act guidelines and calls of other regulatory frameworks. Another ethical concern is accountability. Since OpenAI has disclaimed that any of ChatGPT's responses cannot play as clinical advice, they will not be accountable for any false or misleading responses leading to severe clinical accidents. Accountability thus rests with deploying organizations to establish liability frameworks and indemnification policies. Any organization must present high organizational readiness in using AI and establishing a comprehensive framework for technical training, technology oversight, accountability chains and also the compliance with the regulatory framworks (Kruszyńska-Fischbach et al., 2022).

## 5.5 Potential Solution - Technological

Our study demonstrated that LLMs can have strong preliminary performance in controlled simulations in delivering high-quality mental health support, though real-world efficacy remains untested. However, randomness and instability in the responses also emphasized the immaturity in directly using ChatGPT in mental health interventions, particularly for users with clinical symptoms, highlighting both stability and validity gaps, as well as potential data-privacy concerns in live deployments. Several technological approaches can mitigate some risks; these should be validated via pilot deployments with clinician feedback. A multi-agent model can be designed to bypass the limitations of each AI application, where a validation agent scores responses against predefined empathy and safety benchmarks (Guo et al., 2024). This process is similar to that in our study, as we could evaluate whether a response satisfies our expectations of positivity and empathy, and psychologists can define in-depth criteria to ask AI to measure. If the response fails to satisfy our predefined benchmark, it will ask the counseling AI agent to regenerate (Gronauer & Diepold, 2022; Mökander et al., 2023). Another AI agent could also be designed to evaluate a client's concurrent risk level, triggering automatic crisis protocols (e.g., suicide-risk flags) and human escalation. If it identified a client presenting a risk level higher than the benchmark, it could automatically refer the client's information to human counselors for immediate clinical intervention. Meanwhile, refining prompts emphasizing the avoidance of unqualified answers might also help. Along with the continuous advancement in technology and human-computer interactions, various approaches can be developed to further improve the real-life performance of AI in mental health care.

*5.6 Potential Solution - Applicational*

Considering its current capabilities and limitations, ChatGPT and similar chatbots are best used for general mental health support and prevention rather than high-risk clinical interventions. For example, in the school context, it has been proposed that schools first make AI applications available for students without any diagnosed mental disorders, allowing them to largely contribute to students' overall well-being. Effective early interventions could foreseeably prevent the occurrence of mental health disorders among students, with clear informed-consent procedures and parental notification. All applications are best piloted in non-clinical settings with A/B trials to measure user satisfaction and safety before any clinical rollout.

The combination of AI chatbots and mobile health applications with wearable devices could provide opportunities for continuous monitoring and real-time personalized feedback, thus improving preventive care (Knight & Bidargaddi, 2018). Utilizing these low-cost and widely applied AI applications, mental health prevention can be conducted at a large organizational or societal level and designed to adapt to a specific population's needs, such as customized AI chatbots for young people. If AI applications could reduce depression and anxiety symptoms, the number of patients requiring traditional mental health facilities could be reduced, thus contributing to global mental health care, though such system-wide impacts remain hypothetical until confirmed by large-scale trials.

*5.7 Potential Solution - Organizational*

Besides paying attention to the use of AI in mental health support and intervention, organizations such as schools and colleges can focus on overseeing the service procedure and quality assurance during the process to improve the effectiveness of mental health intervention and the utilization efficiency of AI. By using protocols for continuous review and assessment of AI responses, human supervisors can timely correct any inaccuracies or inconsistencies that might arise. In this way, the AI can remain built to standards of empathy, warmth, and acceptance. In sum, human oversight can facilitate the integration of AI tools with traditional counseling services, creating a hybrid model that leverages the strengths of both. If organizations would like to utilize AI in mental health intervention, it is crucial to understand its limitations and create a product roadmap, risk-management plans, financial budgeting, and human resources training, to make the AI services blend into the traditional psychological services and ethical safeguards.

Supplementary Dataset S1. Full result matrix for sentiment and empathy metrics across 80 prompts. Source: Authors' own work.

**Author Contributions:** Conceptualization, Y.N.; methodology, Y.N., Y.C.; software, Y.C.; validation, Y.N.; formal analysis, Y.C., Y.N.; investigation, Y.N.; resources, Y.N.; data curation, Y.N., Y.C.; writing—original draft

preparation, Y.N., Y.C.; writing—review and editing, Y.N.; visualization, Y.C., Y.N.; project administration, Y.N. All authors have read and agreed to the published version of the manuscript.

**Funding:** This research received no external funding.

**Institutional Review Board Statement:** Not applicable.

**Informed Consent Statement:** Not applicable

**Conflicts of Interest:** The authors declare no conflicts of interest.

**References**

1. Adeshola, I. & Adepoju, A. P. (2023) 'The opportunities and challenges of ChatGPT in education', *Interactive Learning Environments*, 0(0), pp. 1–14. doi:10.1080/10494820.2023.2253858.

2. American Psychological Association (2013) 'The therapist effect', *APA Monitor on Psychology*, February. Available at: https://www.apa.org/monitor/2013/02/therapist (Accessed 02 Jul 2024).

3. American Psychological Association (2017) 'APA PsycNET record no. 2017-19174-002', *APA PsycNET*. Available at: https://psycnet.apa.org/record/2017-19174-002 (Accessed 30 Apr 2025).

4. Areán, P. A. (2021) 'Here to stay: digital mental health in a post-pandemic world…', *Technology, Mind, and Behavior*, 2, e00073.

5. Ayala, M. (2023) 'The drawbacks and downsides of online therapy, internet counseling', *Business Insider*, Apr. Available at: https://www.businessinsider.com/online-therapy-… (Accessed 01 May 2025).

6. BBC News (2024) 'AI and mental health: Young people turning to AI therapist bots', *BBC News*, 9 Jan. Available at: https://www.bbc.com/news/technology-67872693 (Accessed 01 May 2025).

7. Biswas, S. S. (2023) 'Role of Chat GPT in public health', *Annals of Biomedical Engineering*, 51(5), pp. 868–869. doi:10.1007/s10439-023-03172-7.

8. Blei, D. M., Ng, A. Y. & Jordan, M. I. (2003) 'Latent Dirichlet allocation', *Journal of Machine Learning Research*, 3, pp. 993–1022.

9. BMJ Group (2024) 'Don't rely on AI chatbots for accurate, safe drug information, patients warned', *BMJ Quality & Safety* [Press release].

10. Brown, T. B. et al. (2020) 'Language models are few-shot learners', *arXiv*. doi:10.48550/arXiv.2005.14165.

11. Castonguay, L. G. & Hill, C. E. (eds.) (2012) *Transformation in Psychotherapy: Corrective Experiences Across Cognitive Behavioral, Humanistic, and Psychodynamic Approaches*. Washington, DC: American Psychological Association.

12  Caughey, J. L. (1984) *Imaginary Social Worlds: A Cultural Approach to Media-Cultivated Relationships*. Lincoln: University of Nebraska Press.

13  Chen, M. et al. (2022) 'Negative information measurement at AI edge…', *ACM Transactions on Internet Technology*, 22, pp. 1–16. doi:10.1145/3471902.

14  Chervenak, J. et al. (2023) 'The promise and peril of using a large language model to obtain clinical information…', *Fertility and Sterility*. doi:10.1016/j.fertnstert.2023.05.151.

15  Coombs, N. C., Meriwether, W. E., Caringi, J. & Newcomer, S. R. (2021) 'Barriers to healthcare access among U.S. adults with mental health challenges', *SSM Population Health*, 15, 100847.

16  Danieli, M. et al. (2022) 'Assessing the impact of conversational artificial intelligence in the treatment of stress and anxiety in aging adults', *JMIR Mental Health*, 9(9), e38067. doi:10.2196/38067.

17  D'Alfonso, S. (2020) 'AI in mental health', *Current Opinion in Psychology*, 36, pp. 112–117. doi:10.1016/j.copsyc.2020.04.005.

18  Dergaa, I. et al. (2024) 'ChatGPT is not ready yet for use in providing mental health assessment and interventions', *Frontiers in Psychiatry*, 14. doi:10.3389/fpsyt.2023.1277756.

19  Eisenberg, D., Hunt, J. & Speer, N. (2013) 'Mental health in American colleges and universities…', *Journal of Nervous and Mental Disease*, 201(1), pp. 60–67.

20  Elyoseph, Z., Hadar-Shoval, D., Asraf, K. & Lvovsky, M. (2023) 'ChatGPT outperforms humans in emotional awareness evaluations', *Frontiers in Psychology*, 14, 1199058.

21  Farhat, F. (2024) 'ChatGPT as a complementary mental health resource: a boon or a bane', *Annals of Biomedical Engineering*, 52(5), pp. 1111–1114.

22  Kim, B. & Mihaf, M. (2024) 'Her teenage son killed himself after talking to a chatbot. Now she's suing.', *The Washington Post*, 24 Oct., Available at: https://www.washingtonpost.com/nation/2024/10/24/character-ai-lawsuit-suicide/, viewed 2 May 2025.

23  Friis-Healy, E. A., Nagy, G. A. & Kollins, S. H. (2021) 'It is time to REACT…', *JMIR Mental Health*, 8, e25456.

24  Fulmer, R. et al. (2018) 'Using psychological artificial intelligence (Tess)…', *JMIR Mental Health*, 5(4), e64. doi:10.2196/mental.9782.

25  Guo, T. et al. (2024) 'Large language model based multi-agents: a survey of progress and challenges', *arXiv*.

26  Gronauer, S. & Diepold, K. (2022) 'Multi-agent deep reinforcement learning: a survey', *Artificial Intelligence Review*, 55(2), pp. 895–943.

27  Hadar-Shoval, D., Elyoseph, Z. & Lvovsky, M. (2023) 'The plasticity of ChatGPT's mentalizing abilities…', *Frontiers in Psychiatry*, 14.

28  Hamdoun, S. et al. (2023) 'AI-based and digital mental health apps: balancing need and risk', *IEEE Technology and Society Magazine*, 42(1), pp. 25–36.

29 Henson, P., Wisniewski, H., Hollis, C., Keshavan, M. & Torous, J. (2019) 'Digital mental-health apps and the therapeutic alliance: initial review', *BJPsych Open*, 5, e15.

30 Heston, T. F. (2023) 'Safety of large language models in addressing depression', *Cureus*. doi:10.7759/cureus.50729.

31 Hidaka, B. H. (2012) 'Depression as a disease of modernity: explanations for increasing prevalence', *Journal of Affective Disorders*, 140(3), pp. 205–214.

32 Horton, D. & Wohl, R. R. (1956) 'Mass communication and para-social interaction', *Psychiatry*, 19, pp. 215–229.

33 Huang, Q., Zhang, Y. & Lee, M. (2024) 'Benchmarking empathic expression in large language models', *ICWSM*, 18(1), e31338.

34 Hutto, C. & Gilbert, E. (2014) 'VADER: a parsimonious rule-based model for sentiment analysis of social media text', *Proc. ICWSM*, 8(1), pp. 216–225.

35 Joshi, K. (2023) 'AI mental health therapist chatbot', *International Journal for Research in Applied Science and Engineering Technology*.

36 Kamath, R., Ghoshal, A., Eswaran, S. & Honnavalli, P. (2022) 'An enhanced context-based emotion detection model using RoBERTa', *Proc. IEEE CONECCT 2022*, p. 6. doi:10.1109/CONECCT55679.2022.9865796.

37 Kapoor, A. & Goel, S. (2022) 'Applications of conversational AI in mental health: a survey', *Proc. ICOEI 2022*, pp. 1013–1016.

38 Kim, T. & Vossen, P. (2021) 'EmoBERTa: speaker-aware emotion recognition in conversation with RoBERTa', *arXiv*. doi:10.48550/arXiv.2108.12009.

39 Kirch, W. (ed.) (2008) 'Pearson's correlation coefficient', in *Encyclopedia of Public Health*, Dordrecht: Springer, pp. 1090–1091.

40 Knight, A. & Bidargaddi, N. (2018) 'Commonly available activity tracker apps…', *Journal of Affective Disorders*, 236, pp. 31–36.

41 Koopman, B. et al. (2023) 'Dr ChatGPT tell me what I want to hear: how different prompts impact health answer correctness', *Proc. EMNLP 2023*. doi:10.18653/v1/2023.emnlp-main.928.

42 Kruszyńska-Fischbach, A. et al. (2022) 'Organizational e-health readiness…', *International Journal of Environmental Research and Public Health*, 19(7).

43 Lazer, D. et al. (2009) 'Computational social science', *Science*, 323, pp. 721–723.

44 Lattie, E. G., Stiles-Shields, C. & Graham, A. K. (2022) 'An overview of and recommendations for more accessible digital mental-health services', *Nature Reviews Psychology*, 1, pp. 87–100.

45 Levkovich, I. & Elyoseph, Z. (2023) 'Identifying depression and its determinants upon initiating treatment: ChatGPT versus primary care physicians', *Family Medicine and Community Health*, 11(4), e002391.

46  Liu, J. M. et al. (2023) 'ChatCounselor: a large language model for mental health support', *arXiv*. doi:10.48550/arXiv.2309.15461.

47  Manovich, L. (2017) 'Cultural analytics: computational approaches to the study of culture', in Schäfer, M. & van Es, K. (eds) *The Datafied Society*, Amsterdam University Press, pp. 55–68.

48  Maulana, F. I. et al. (2024) 'Sentiment analysis on service quality of an online healthcare mobile platform…', in Triwiyanto, T., Rizal, A. & Caesarendra, W. (eds) *Proc. 4th Int. Conf. Electronics, Biomedical Engineering & Health Informatics*, Springer, pp. 383–394.

49  Maurya, R. K. (2023) 'A qualitative content analysis of ChatGPT's client simulation role-play…', *Counseling and Psychotherapy Research*. doi:10.1002/capr.12699.

50  May, N. (2024) '"Alarm bells should be going off" as mental health counseling app expands into Australia, critics say', *The Guardian*, 17 Jun.

51  Moilanen, J. et al. (2022) 'Measuring the effect of mental health chatbot personality on user engagement', *Proc. MUM 2022*, Lisbon, pp. 138–150.

52  Mökander, J., Schuett, J., Kirk, H. R. & Floridi, L. (2023) 'Auditing large language models: a three-layered approach', *AI Ethics*. doi:10.1007/s43681-023-00289-2.

53  Morris, R. R., Kouddous, K., Kshirsagar, R. & Schueller, S. M. (2018) 'Towards an artificially empathic conversational agent for mental health applications', *Journal of Medical Internet Research*, 20(6), e10148.

54  Moulya, S. & Pragathi, T. R. (2022) 'Mental health assist and diagnosis conversational interface…', *Journal of Physics: Conference Series*, 2161(1), 012039.

55  Ni, Y., & Jia, F. (2025). A Scoping Review of AI-Driven Digital Interventions in Mental Health Care: Mapping Applications Across Screening, Support, Monitoring, Prevention, and Clinical Education. *Healthcare*, 13(10), 1205. https://doi.org/10.3390/healthcare13101205

56  Nichele, E., Lavorgna, A. & Middleton, S. (2022) 'Identifying key challenges and needs in digital mental health moderation practices…', *SN Social Sciences*, 2.

57  Pang, B. & Lee, L. (2008) 'Opinion mining and sentiment analysis', *Foundations and Trends in Information Retrieval*, 2(1–2), pp. 1–135.

58  Pascoe, M. C., Hetrick, S. E. & Parker, A. G. (2020) 'The impact of stress on students in secondary school and higher education', *International Journal of Adolescence and Youth*.

59  Portt, E., Person, S., Person, B., Rawana, E. & Brownlee, K. (2020) 'Empathy and positive aspects of adolescent peer relationships', *Journal of Child and Family Studies*, 29(9), pp. 2416–2433.

60  Prescott, M. R. et al. (2022) 'Real-world effectiveness of digital mental-health services during the COVID-19 pandemic', *PLoS ONE*, 17, e0272162.

61  Proferes, N., Jones, N., Gilbert, S., Fiesler, C. & Zimmer, M. (2021) 'Studying Reddit: a systematic overview of disciplines, approaches, methods, and ethics', *Social Media + Society*, 7(2), 1–14.

62 Rathnayaka, P. et al. (2022) 'A mental health chatbot with cognitive skills for personalised behavioural activation…', *Sensors*, 22(10), 3653.

63 Reeves, B. & Nass, C. (1996) *The Media Equation*. Cambridge: Cambridge University Press.

64 Rollwage, M., Juchems, K., Habicht, J., Carrington, B., Hauser, T. & Harper, R. (2022) 'Conversational AI facilitates mental-health assessments and is associated with improved recovery rates', *medRxiv*. doi:10.1101/2022.11.03.22281887.

65 Rollwage, M., Habicht, J., Juechems, K., Carrington, B., Stylianou, M., Hauser, T. U. & Harper, R. (2023) 'Using conversational AI to facilitate mental health assessments and improve clinical efficiency…', *JMIR AI*, 2, e44358.

66 Roose, K. (2024) 'How Claude Became Tech Insiders' Chatbot of Choice', *The New York Times*, 13 Dec.

67 Roumeliotis, K. I. & Tselikas, N. D. (2023) 'ChatGPT and Open-AI models: a preliminary review', *Future Internet*, 15, 200.

68 Rubin, A. M., Perse, E. M. & Powell, R. A. (1985) 'Loneliness, parasocial interaction, and local television news viewing', *Human Communication Research*, 12, pp. 155–180.

69 Rubin, M., Arnon, H., Huppert, J. D. & Perry, A. (2024) 'Considering the role of human empathy in AI-driven therapy', *JMIR Mental Health*, 11(1).

70 Sallam, M., Egger, J., Roehrig, R. & Puladi, B. (2024) 'Navigating the Peril of Generated Alternative Facts: A ChatGPT-4 fabricated Omega variant case…', *arXiv* 2403.09674.

71 Sass, K., Fetz, K., Oetken, S., Habel, U. & Heim, S. (2013) 'Emotional verbal fluency: a new task on emotion and executive function interaction', *Behavioral Sciences*, 3(3), pp. 372–387.

72 Secinaro, S., Calandra, D., Secinaro, A., Muthurangu, V. & Biancone, P. (2021) 'The role of artificial intelligence in healthcare: a structured literature review', *BMC Medical Informatics and Decision Making*, 21(1), 125.

73 Shaik, T. et al. (2022) 'AI enabled RPM for mental health facility', *Proc. ACM Workshop on Mobile & Wireless Sensing for Smart Healthcare*, pp. 26–32.

74 Shao, R. (2023) 'An empathetic AI for mental health intervention: conceptualizing and examining artificial empathy', *2nd Empathy-Centric Design Workshop*, pp. 1–6.

75 Shah, F., Singh, R. P. & Satish, D. (2023) 'A Keras functional conversational AI agent for psychological condition analysis', *IJRASET*, 11(2), pp. 1171–1177.

76 Sharma, A., Miner, A. S., Atkins, D. C. & Althoff, T. (2020) 'A computational approach to understanding empathy expressed in text-based mental health support', *arXiv*.

77 Sharma, A. et al. (2022) 'Human-AI collaboration enables more empathic conversations in text-based peer-to-peer mental health support', *arXiv*.

78 Shrout, P. E. & Fleiss, J. L. (1979) 'Intraclass correlations: uses in assessing rater reliability', *Psychological Bulletin*, 86(2), pp. 420–428.

79 Su, S. et al. (2022) 'Efficacy of artificial intelligence-assisted psychotherapy in patients with anxiety disorders', *Frontiers in Psychiatry*, 12.

80 Thieme, A. et al. (2023) 'Designing human-centered AI for mental health: developing clinically relevant applications…', *ACM Transactions on Computer-Human Interaction*, 30(2), 1–50.

81 Trappey, A. J. C. et al. (2022) 'Development of an empathy-centric counseling chatbot system…', *Processes*, 10(5), 930.

82 Tutun, S. et al. (2023) 'An AI-based decision support system for predicting mental health disorders', *Information Systems Frontiers*, 25(3), pp. 1261–1276.

83 Vahedifard, F., Haghighi, A. S., Dave, T., Tolouei, M. & Zare, F. H. (2023) 'Practical use of ChatGPT in psychiatry for treatment plan and psychoeducation', *Semantic Scholar* (preprint).

84 Vilaza, G. N. & McCashin, D. (2021) 'Is the automation of digital mental health ethical?', *Frontiers in Digital Health*, 3, 689736.

85 Wampold, B. E., Baldwin, S. A., Holtforth, M. G. & Imel, Z. E. (2017) 'What characterizes effective therapists?', in Castonguay, L. G. & Hill, C. E. (eds) *How and why are some therapists better than others?* Washington, DC: APA, pp. 37–53.

86 Wang, L., Suarez-Ibarrola, R. & Gupta, S. (2023) 'Context length, hallucination and stability in ChatGPT-4', *arXiv*.

87 Wu, T. et al. (2023) 'A brief overview of ChatGPT: history, status-quo and potential future development', *IEEE/CAA Journal of Automatica Sinica*, 10, pp. 1122–1136.

88 Wrightson-Hester, A.-R. et al. (2023) 'An artificial therapist (Manage Your Life Online) to support the mental health of youth', *JMIR Human Factors*, 10, e46849.

89 Yalçın, Ö. N. & DiPaola, S. (2020) 'Modeling empathy: building a link between affective and cognitive processes', *Artificial Intelligence Review*, 53(4), pp. 2983–3006.